\RequirePackage[2020-02-02]{latexrelease}

\documentclass[aps,preprintnumbers,showkeys,showpacs]{revtex4}
\usepackage{epsfig}
\usepackage{psfrag}
\usepackage{amsfonts}
\usepackage{graphicx}
\usepackage{dcolumn}
\usepackage{bm}
\usepackage{setspace}
\usepackage{multirow}
\usepackage{amsmath}
\usepackage{slashbox}

\graphicspath{{pic/}}

\begin{document}
\title{Probing solar modulation analytic models with cosmic ray periodic spectra}
\author{Wei-Cheng Long}
\author{Juan Wu}
\email{wu@cug.edu.cn}
\affiliation{School of Mathematics and Physics, China University of Geosciences, Wuhan 430074, China}

\begin{abstract}
The AMS02 experiment has published the periodic spectra of proton, helium and helium isotopes across the majority of the 24 solar cycle. These precise data exhibit temporal structures that correlate with solar modulation. In this study, we utilize these data to probe three analytic solar modulation models, including the force-field approximation, the convection-diffusion model and the extended force-field approximation with a drift effect. We adopt a method that eliminates the influence of interstellar cosmic ray spectra, and use the Earth-observed spectra at time $t_1$ to predict those at time $t_2$. In order to explore the rigidity-dependence of solar modulation models, we substitute the conventional potential parameter $\phi$ with a modified parameter $\phi'=\frac{R}{ k_2(R)}\phi$ for our analysis. Combining with the $\chi^2$ minimization method, the best-fit modulation parameter $\phi'$ can be evaluated. First, we test the validity of a rigidity-independent $\phi'$ and find that both the force-field approximation (FFA) and the extended force-field approximation (EFFA) agree well with data near the solar minimum period. However, all models significantly deviate from the data during the solar maximum. Consequently, we assume a constant $\phi'(t_1)$ at solar minimum and calculate $\Delta\phi'=\phi'(t_2)-\phi'(t_1)$ for each rigidity bin at time $t_2$. It is found that $\Delta\phi'$ generally adheres to a linear-logarithm relationship with rigidity at any given time. By adopting a linear-logarithm formula of $\Delta\phi'$, we further discover that both the modified FFA and EFFA can reconcile the observations during solar maxima. This suggests that at solar maximum, the parameter $\phi'$, which correlates with the diffusion pattern in the heliospheric magnetic fields, exhibits a rigidity dependence. Moreover, the modified EFFA enhances the concordance with data during periods of pronounced dips as observed by AMS02. This implies that the drift effect could significantly contribute to these solar transient phenomena.
\end{abstract}


\maketitle
\section{Introduction} \label{sec:intro}
Due to the interaction with the heliospheric magnetic fields (HMF) embedded in the solar wind \citep{parkerCosmicRayModulationSolar1958}, the cosmic ray (CR) energy spectra detected at the top of the atmosphere(TOA) of Earth differ from those at the local interstellar space (LIS). The entire process, which CRs undergo within the solar system, is referred to as solar modulation. During the passages of CR particles traversing through the interplanetary space, the modulation effects for them include convection driven by the solar wind, diffusion induced by the small-scale magnetic field irregularities, drift occurring in the large-scale magnetic structures and adiabatic losses due to the expansion of the solar wind \citep{potgieterSolarModulationCosmic2013}.

Solar modulation effects are inherently influenced by the solar activities. The number and surface area of sunspots relate with the intensity of solar activity, and observations on them show a 11-year solar cycle. Meanwhile, the solar magnetic field undergoes a polarity reversal at the solar maximum, suggesting a 22-year cycle for solar activity~\citep{hathawaySolarCycle2015,baloghHeliosphereSolarActivity2008}. These periodic variations in solar activities causes both the solar modulation and the cosmic ray energy spectra, which are affected by solar modulation, to change periodically over time. Direct observation experiments such as PAMELA and AMS-02 have performed long-term measurements on CRs~\citep{adrianiTIMEDEPENDENCEPROTON2013, martucciProtonFluxesMeasured2018, marcelliHeliumFluxesMeasured2022, amscollaborationObservationFineTime2018, aguilarPropertiesCosmicHelium2019a, amscollaborationPeriodicitiesDailyProton2021, amscollaborationPropertiesDailyHelium2022}. The unprecedented accurate periodic data reveal that the CR intensities display time structures that are anti-correlated with solar activities. These data offer substantial potential for us to reveal the features of solar modulation.

The CRs' transport processes in the heliosphere is usually described by Parker equation~\cite{parkerPassageEnergeticCharged1965}. It can be solved either through numerical methods or analytical methods. While the numerical models provide more accurate and physically reasonable solutions \citep{kapplChargesignDependentSolar2016,Vittino:2017Nc,boschiniPropagationCosmicRays2018,cortiNumericalModelingGalactic2019}, they necessitate a comprehensive understanding on the details of various physical quantities in the heliosphere, coupled with strong computational power. In contrast, analytical methods, which rely on a set of simple assumptions, may yield less precise results. However, these methods are computationally efficient and hence frequently employed by researchers. The most commonly used analytic models are the force-field approximation (FFA) and the convection-diffusion model (CD)~\citep{moraalCosmicRayModulationEquations2013}. These models describe solar modulation by using a single parameter and greatly enhance the convenience of their applicability~\citep{caballero-lopezLimitationsForceField2004,engelbrechtUncertaintiesImplicitUse2020}. A robust solar modulation model is crucial for comprehending CR acceleration and propagation~\citep{boschiniSolutionHeliosphericPropagation2017b, tomassettiSolarNuclearPhysics2017, wuRevisitCosmicRay2019, wangTestingConsistencyProapgation2022, maurinHaloSizeBe2022} as well as for detecting dark matter signal~\citep{ellisParticlesCosmologyLearning2000, yuanSystematicStudyUncertainties2015, chengAMS02PositronExcess2017, zhuDarkMatterAntiproton2022}.

As we known, the modulation effects might be very sensitive to the particle's rigidity\citep{gleesonEnergyLossesModulation1971}. For a CR particle, the diffusion coefficient is $\kappa=v \lambda / 3$, in which $v$ is the particle's velocity and $\lambda$ is its mean free path. This coefficient is important for us to understand the scattering of particles on the random heliospheric magnetohydrodynamic waves and discontinuities. When the particle resonates with the HMF fluctuations with a spectrum $w(k)\sim k^{-\eta}$, where $k$ is the resonant wave number, the diffusion mean free path can be expressed as $\lambda \sim R^{2-\eta}$~\citep{jokipiiCosmicRayPropagationCharged1966,jokipiiCosmicRayPropagationII1967,jokipiiPropagationCosmicRays1971,engelbrechtTheoryCosmicRay2022,tomassettiDataDrivenAnalysis2023}. 
A fundamental prerequisite for FFA is that the mean free path $\lambda$ of a particle is proportional to its rigidity, i.e., $\eta=1$. 
%under which data could be reconciled in the vicinity of solar minimum with FFA as well as CD \citep{gloecklerLowEnergyCosmicRayModulation1966, jokipiiCosmicRayPropagationII1967,ogallagherHeliocentricIntensityGradients1967}. 
However, the real relationship between $\lambda$ and the rigidity remains uncertain. If $\eta\neq 1$, the modulation parameter may have a different rigidity dependence. In recent years, there has been a lot of research focusing on the modification of FFA. Some studies have investigated the rigidity dependence of FFA and have updated FFA analytic formula by introducing additional parameters~\citep{cortiSOLARMODULATIONLOCAL2016a, gieselerEmpiricalModificationForce2017,shenSolarModulationGalactic2021}. 
%They didn't give a theoretical deduction between the  rigorous deduction and discussion on the role of diffusion coefficient.
There are also some studies \cite{cholisPredictiveAnalyticModel2016, kuhlenTimeDependentAMS02ElectronPositron2019} on FFA have considered a drift effect, which is prominent near the solar minimum and negligible during the HMF polarity reversal period~\citep{aslamUnfoldingDriftEffects2023}. The inclusion of this effect extend FFA into a charge-sign dependent model. 

However, these studies usually firstly assumed CR LIS spectra, and then combined the experimental data to constrain the modulation parameters. It is known that CR LIS spectra have only been measured by Voyager-1 below a few hundred MeV~\cite{Cummings2016}. Above this energy, no experimental data exist, leading to the adoption of different LIS models in the literature. Therefore, the analysis results might be biased due to inaccurate assumptions of LIS spectra. To get rid of the impact of LIS energy spectra, an alternative method (herein referred to as the Non-LIS method here) was proposed in~\cite{cortiTestValidityForcefield2019}. In their work, they rewrote the FFA formula by removing the term of LIS spectra. Instead, they calculated the CR TOA intensity at time $t_2$ ($J(t_2)$) based on the TOA intensity at $t_1$ ($J(t_1)$). This method does not require an assumption of LIS spectra, but only needs to use the periodic CR data for the analysis. An important finding from their work was that the validity of FFA varies at different periods of solar activity. In this work, we will utilize the Non-LIS method to further investigate the rigidity dependence of the solar modulation effect for various analytic modulation models, including FFA, CD and an extended FFA (with a drift effect). 

\section{Description of analytic solar modulation models}
\label{sec:semi}
\subsection{Force-field approximation}
The basic transport equation(TPE) was first derived by Parker in the solar wind reference frame~\citep{gleesonEnergyChangesCosmic1978, caballero-lopezLimitationsForceField2004,moraalCosmicRayModulationEquations2013}:
\begin{equation}\label{eq:tpe3}
\frac{\partial f}{\partial t} + \nabla \cdot(\mathbf Vf - \mathbf K\cdot \nabla f) -\frac{1}{3p^2}(\nabla\cdot \mathbf V)\frac{\partial f}{\partial  p}(p^3 f) =Q,
\end{equation}
where $f(\mathbf r, p, t)$ is the phase space density or the omini-directional distribution function of CRs as a function of position $\mathbf r$, momentum $p$ and time $t$. It is linked to the differential intensity in terms of energy by: $J_T = p^2 f\left( \mathbf r, p, t \right)$. In Eq.~\eqref{eq:tpe3}, $\mathbf V$ is the solar wind velocity, $ \mathbf K$ is the diffusion tensor and $Q$ is the local sources in the heliosphere. The diffusion tensor can be expressed as $\mathbf K = \mathbf K_a + \mathbf K_s$, in which $\mathbf K_s$ denotes the symmetrical component of the diffusion tensor and $\mathbf K_a$ represents the asymmetrical component associated with the drift effect.

After that, an equivalent equation was derived in the observer's reference frame~\citep{gleesonCosmicRaysInterplanetary1967}: 
\begin{equation}  \label{eq:tpe2}
\frac{\partial f}{\partial t} + \nabla \cdot (C\mathbf Vf - \mathbf K\cdot \mathbf{\nabla} f)+\frac{1}{p^2}\frac{\partial}{\partial p}(p^2\langle\dot p\rangle f) = Q. 
\end{equation}
Here $C = -(1/3)\partial \ln f/\partial \ln p$ is the Compton-getting coefficient, which corrects the anisotropy of transformation from the wind reference frame to the stationary reference frame~\citep{gleesonComptongettingEffect1968}. 

FFA solves Eq.~\eqref{eq:tpe2} under a series of assumptions: a) no local source of CRs, i.e. $Q=0$; b) the existence of a steady state, i.e. $\partial f/\partial t = 0$; c) spherically symmetric and ignoring the drift effect; d) an adiabatic rate $\langle \dot{p}\rangle = (p/3)\mathbf V\cdot \nabla f/f = 0$. Additionally, the streaming term $(C\mathbf Vf - \mathbf K\cdot \mathbf{\nabla} f)$ is assumed to be divergence free and only the radial component of $\mathbf K$, denoted as $\kappa$ for convenience, is nonzero. Consequently, Eq. \eqref{eq:tpe2} can be written as the so-called force-field equation: 
\begin{equation} \label{eq:ffaeq}
% C\mathbf Vf - \kappa\frac{\partial f}{\partial r} = 0
\frac{\partial f}{\partial r} + \frac{Vp}{3\kappa}\frac{\partial f}{\partial p} = 0,
\end{equation}
where $\kappa$ is the one-dimension diffusion coefficient, $r$ is the radial position and $V$ is the radial component of solar wind velocity.

It is a first-order partial differential equation with a characteristic equation 
\begin{equation}\label{eq:ffacharacter}
    \frac{\mathrm d p}{\mathrm d r} = \frac{pV}{3\kappa}. 
\end{equation}

The solution of Eq.~\eqref{eq:ffaeq} is 
\begin{equation}
f\left(r_{TOA}, p_{TOA} \right)  = f\left( r_{LIS}, p_{LIS} \right),
\end{equation}
where $r_{LIS}$, $p_{LIS}$ and $r_{TOA}$, $p_{TOA}$ are the positions and momenta of CR particles in LIS and at TOA, respectively, and $f(r,~p)$ is the density of CRs along the characteristic curve described by Eq.~\eqref{eq:ffacharacter}.

Assuming the diffusion coefficient $\kappa$ could be separated in radius distance $r$ and rigidity $R\equiv pc/Ze$, i.e. $\kappa(r, R) = \beta k_1(r) k_2(R)$, where $Z$ is the charge number of the particle, $\beta$ is the ratio of the particle velocity $v$ and the speed of light $c$, we can define the modulation parameter $\phi$ as
\begin{equation}\label{eq:phi}
    \int_{R_{TOA}}^{R_{LIS}}\frac{ k_2(R)\beta}{R} \mathrm d R= \int_{r_{TOA}}^{r_{LIS}} \frac{V}{3 k_1(r)}\mathrm dr \equiv \phi,
\end{equation}
where $R_{LIS}$ and $R_{TOA}$ are the rigidities of CR particles in LIS and at TOA, respectively. Both $\phi$ and $k_2(R)$ are independent with the CR species but vary with time. Assuming that $T_{LIS}$ and $T_{TOA}$ represent the energies of a CR particle in LIS and at TOA, the energy loss experienced by this particle in the heliosphere can be denoted as $\Phi\equiv T_{LIS}-T_{TOA}$. If $\Phi \ll E_0$, where $E_0$ is the stationary energy of the particle, then the relationship between $\Phi$ and $\phi$ becomes~\citep{gleesonSolarModulationGalactic1968}
\begin{equation}\label{eq:Phiphi}
    \Phi(R) =|Z|e \frac{R}{ k_2(R)} \phi.
\end{equation}

From Eq. \eqref{eq:Phiphi}, we can define the modified modulation potential $\phi'$ as
\begin{equation}\label{eq:modifiedphi}
    \phi' \equiv \frac{\Phi}{|Z|e} =\frac{R}{ k_2(R)}\phi.
\end{equation}

Taking $\lambda\propto R$~\citep{gloecklerLowEnergyCosmicRayModulation1966,tomassettiTestingDiffusionCosmic2018}, thus $k_2(R) = R$. From the definition of Eq. \eqref{eq:phi}, it can be found that $\phi$ is a rigidity-independent parameter. So $\Phi = Ze\phi$ for any arbitrary rigidity. This leads to the conventional FFA, which condenses all the physical processes into a single parameter $\phi$ (or $\Phi$)~\citep{gleesonSolarModulationGalactic1968,gleesonStudyForcefieldEquation1973}. 
It should be noticed that if $ k_2(R) = R$ is not the condition, we may get more general results by substituting $\phi'$ for $\phi$. In this case, $\phi'$ could become rigidity-dependent, and $\Phi$ can be expressed as $\Phi=Ze\phi'$.

The modulated spectrum $J(r_{TOA},T_{TOA})$ and the unmodulated LIS spectrum $J\left(r_{LIS},T_{TOA}+\Phi\right)$ are related as
\begin{equation} \label{eq:ffacvt}
\begin{aligned}    
    J(r_{TOA},T_{TOA}) =& \frac{T_{TOA}(T_{TOA}+2E_0)}{\left( T_{TOA}+Ze\phi' \right) \left( T_{TOA}+Ze\phi' + 2E_0 \right) }\\
&\times J\left(r_{LIS},T_{TOA}+ Ze\phi'\right).
\end{aligned}
\end{equation}
Therefore, by studying the modified potential parameter $\phi'$, we can investigate the FFA's rigidity dependence effect.

\subsection{Convection-Diffusion equation}
The other analytic modulation model CD can be directly derived from Eq. \eqref{eq:tpe3}. The assumptions made in CD is similar as those given in FFA. Then Eq.~\eqref{eq:tpe3} can be simplified into the CD equation
\begin{equation}\label{eq:cd}
Vf- \kappa \frac{\partial f}{\partial r} = 0, 
\end{equation}
for which the solution is
\begin{equation}\label{eq:cd_solution}
    f\left(r_{TOA}, p_{TOA} \right)  = f\left( r_{LIS}, p_{LIS} \right)e^{-M}, 
\end{equation}
where 
\begin{equation*}\label{eq:cd_M}
\quad M \equiv  \int_{r_{TOA}}^{r_{LIS}}\frac{V}{\kappa}\mathrm d r.
\end{equation*}
According to Eq.~\eqref{eq:phi}, Eq.~\eqref{eq:modifiedphi} and Eq.~\eqref{eq:cd_solution}, the relation between $\phi$ ($\phi'$) and $M$ is \citep{quenbyTheoryCosmicrayModulation1984,moraalCosmicRayModulationEquations2013}
\begin{equation}\label{eq:M}
M = \frac{3\phi}{\beta k_2(R)} = \frac{3\phi'}{\beta R}.
\end{equation}

FFA and CD have been widely used due to their simplicity. Both of them can compress the modulation processes into one single parameter, which depends on the specific form of $k_2(R)$. Unlike FFA, which assumes $ k_2(R) = R$, CD does not place any constraints on $ k_2(R)$. It may allow the modulation effect expected by CD to be rigidity-dependent. 

\subsection{Extended force-field approximation}
Both FFA and CD neglect the drift effect. By incorporating the drift effect, Eq.~\eqref{eq:ffaeq} was led to Kuhlen's extended FFA (EFFA) equation~\cite{kuhlenTimeDependentAMS02ElectronPositron2019}:
\begin{equation}\label{eq:effaeq} 
   \frac{\partial f}{\partial r} + \frac{pV}{3\kappa}\frac{\partial f}{\partial p} = \frac{v_{d,r}}{\kappa}f,
\end{equation}
in which all the quantities are angular averaged, and $v_{d,r}$ is the radical component of the averaged drift velocity.

The solution of Eq.~\eqref{eq:effaeq} is
\begin{equation}\label{eq:kuhlen-solution} 
    \begin{aligned}
        f\left(r_{TOA}, p_{TOA} \right) =& f(r_{LIS}, p_{LIS})%\\
    %&
    \times \exp{\left[ -\int_{{r_{TOA}}}^{{r_{LIS}}} {} \: d{r} {\frac{v_{d,r}(r, p)}{\kappa(r, p)}} \right]}.
    \end{aligned}
\end{equation}
whose characteristics curve is also Eq.\eqref{eq:ffacharacter}.
This solution's form seems to be a combination of FFA and CD, except for that the solar wind velocity in the integral term is replaced by the drift velocity. The rigidity dependence of $v_{d,r}$ correlates with the behavior of the anti-symmetrical diffusion coefficient $\kappa_a$, and is given as~\citep{burgerRigidityDependenceCosmic2000}
\begin{equation}
    v_{d,r} \propto \frac{\beta R }{3 B}\frac{10  R ^2 }{1 + 10 R^2 }, 
\end{equation}
where $B$ is the magnitude of the heliospheric magnetic field on a large scale.

Given the similarity between the spatial integral part in Eq. \eqref{eq:kuhlen-solution} and the definition of $M$ in Eq. \eqref{eq:cd_solution}, when $\Phi \ll E_0$, this integral term can be associated with $\phi'$ as
\begin{equation}
    \int_{{r_{TOA}}}^{r_{LIS}} {} \: d{r} {\frac{v_{d,r}(r,p)}{\kappa(r, p)}} \longrightarrow g \frac{R\phi}{ k_2(R)} \frac{ 10R ^2 }{1 + 10 R^2 } = g \phi' \frac{ 10R ^2 }{1 + 10 R^2 }
,\end{equation}
where $g$ is a scaling factor related to the magnitude of magnetic field $B$ and the solar wind velocity $V$. A larger $g$ indicates a stronger impact on the CR flux caused by the drift effect.

In summary, all three analytical models are derived from equivalent forms of TPE. Both FFA and CD are one-parameter models that disregard a term related to the adiabatic energy. But these terms differ as they are derived in different frames. Other than FFA and CD, Kuhlen's EFFA incorporates a drift effect and depends on two parameters.

\section{Modulation analysis based on the periodic observations}
\subsection{The Non-LIS method}
In order to eliminate the influence of CR LIS energy spectra, we use the Non-LIS method to explore the general properties of above analytical models. Since the nature of the heliospheric diffusion effect remains incompletely understood, we refrain from specifying the formula for $ k_2(R)$. Therefore, we use the modified modulation parameter $\phi'$ to obtain the relationship between $J(t_1)$ and $J(t_2)$. For FFA, assuming $\Delta\phi' = \phi'(t_2) - \phi'(t_1)$, Eq.~\eqref{eq:ffacvt} can be transformed into  
\begin{equation} \label{eq:ffatime}
    \begin{aligned}
        J(r_{TOA}, T_{TOA}, t_2) =& \frac{T_{TOA}(T_{TOA}+2E_0)}{\left( T_{TOA}+Ze\Delta\phi' \right) \left( T_{TOA}+Ze\Delta\phi' + 2E_0 \right) }\\
                            &\qquad\times J\left(r_{TOA},T_{TOA} + Ze\Delta\phi', t_1\right).
    \end{aligned}
\end{equation}

Similar approaches can be applied to CD and EFFA. For CD, we yield
\begin{equation}\label{eq:cd-time}
  J(r_{TOA}, T_{TOA}, t_2) = J(r_{TOA}, T_{TOA}, t_1) \exp{\left(- \frac{3\Delta\phi'}{\beta R} \right)}.
\end{equation}

For EFFA, we get
\begin{equation}\label{eq:kuhlen-time}
 \begin{aligned}
      J&(r_{TOA}, T_{TOA}, t_2) = \frac{T_{TOA}(T_{TOA}+2E_0)}{\left( T_{TOA}+Ze\Delta\phi' \right) \left( T_{TOA}+Ze\Delta\phi' + 2E_0\right) }\\
      &\times J\left(r_{TOA},T_{TOA} + Ze\Delta\phi', t_1\right)\exp{\left(- g  \frac{10 R^2}{1+10R^2}\Delta\phi'\right)}.
 \end{aligned}
\end{equation}

These three models are all written in terms of $\Delta\phi'$. As suggested in~\cite{cortiTestValidityForcefield2019}, traditional FFA may be reliable to describe the solar modulation effect around the solar minimum period. If we select ${t_1}$ near a period of minimal solar activity, it is reasonable to postulate that $\phi'(t_1)=\phi(t_1)$ is rigidity-independent. Therefore, the analysis on parameter $\Delta\phi'$ can characterize the properties of parameter $\phi'(t_2)$. This approach will reveal the characteristics of solar modulation at any given time $t_2$. For CD and EFFA, we also set ${t_1}$ near solar minimum and similarly assume $\phi'(t_1)$ is rigidity-independent. Subsequently, we use the free parameter $\Delta\phi'$ to study modulation effect at other times. Note that in EFFA, except for $\Delta\phi'$, the factor $g$ linked to the drift effect is also allowed to vary freely in the fittings.

\subsection{Test of the rigidity-independent $\Delta\phi'$}

Synodic solar rotation causes the CR flux recurrent variations on the timescale of Bartels Rotations (BRs), which is 27 days for each BR. We use periodic data from AMS02~\cite{amscollaborationObservationFineTime2018}, which provides the measurements of proton (p) flux between 1 GV and 60 GV and helium (He) flux between 1.9 GV and 60 GV from May 2011 to May 2017. It also provides helium-3 ($^{3}$He) flux between 1.9 and 15 GV and helium-4 ($^{4}$He) flux between 2.1 GV and 21 GV from May 2011 to November 2017~\cite{aguilarPropertiesCosmicHelium2019a}. The measurements covered most time of the 24 solar cycle, during which the solar maximum appeared in April 2014, and the HMF polarity reversed from $A<0$ to $A>0$ at that time.

For p and He, BR 2504 (February 18, 2017--March 16, 2017) is selected as $t_1$ in our work, since the measured flux at this time is higher than those at other phases. This indicates that the solar activity is weakest at this time. For $^{3}$He and $^{4}$He, since each of them has been measured in periods of 4 Bartels rotations (108 days), the period from BR 2502 to BR 2505 is selected as $t_1$.
It is assumed that the distribution of flux within rigidity bin $(R_1, R_2)$ follows a power law. Consequently, the flux value at this bin is assigned to the interval center rigidity $R=\sqrt{R_1R_2}$. We employed a cubic spline method to calculate the interpolated flux values for other rigidities within the rigidity range of observation, and utilize a power law distribution to extrapolate flux beyond this range. A least-square analysis using MINUIT package~\cite{jamesMinuitSystemFunction1975} is applied to obtain the $\Delta \phi'$ values for each model, as well as the $\chi^2 / d.o.f$ values at time $t_2$. For EFFA, the best-fit values of parameter $g$ are also estimated. 

\begin{figure*}[htpb!]
    \centering
    \includegraphics[width=0.8\textwidth]{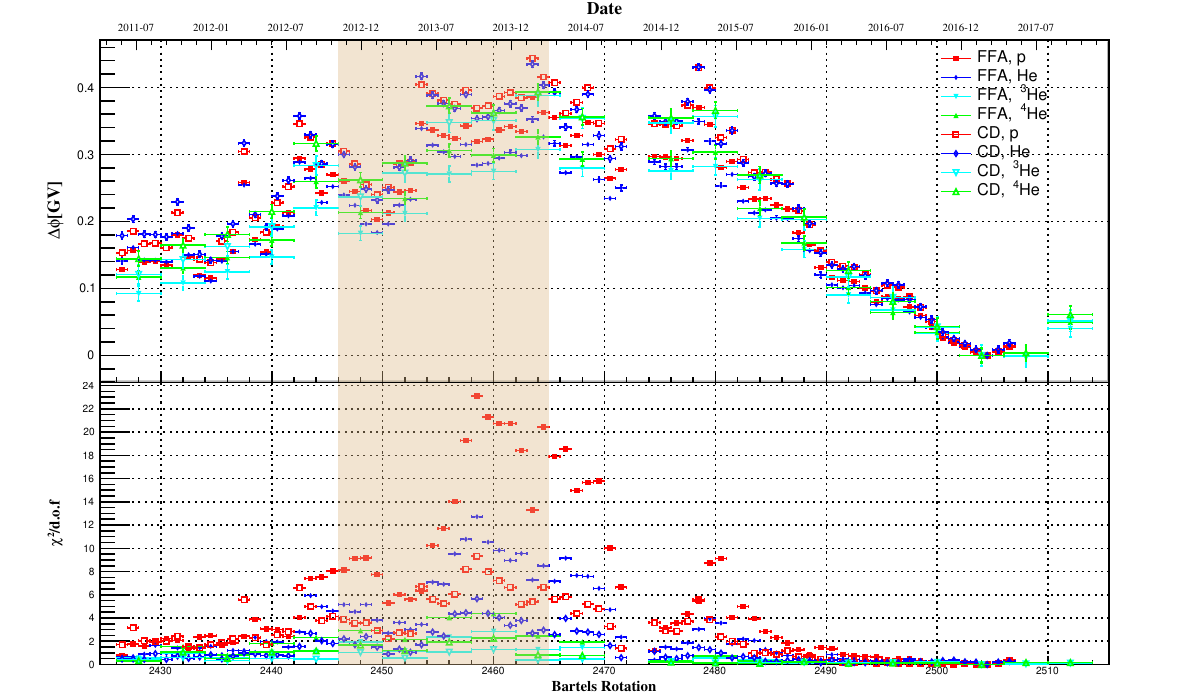}
    \caption{Top panel: the rigidity-independent parameter $\Delta\phi'$ (=$\Delta\phi$) estimated by using the p, He and He isotopes periodic data measured by AMS-02 for FFA and CD model. Bottom panel: the corresponding $\chi^2 / d.o.f$ at each BR. The shade area corresponds to the HMF polarity reversal period from $A < 0$ to $A>0$ (2012 November--2014 March).}
    \label{fig:dphiffacd}
\end{figure*}

First, we use AMS-02 p, He, $^{3}$He and $^{4}$He data to test the validity of a rigidity-independent $\Delta\phi'$ for both FFA and CD. The results are shown in  
Fig.~\ref{fig:dphiffacd}. As we can see, for FFA, at any given time $t_2$, He isotopes yield rather consistent values of $\Delta\phi'$ with p and He. This is also the case for CD. For a given time $t_2$, the best-fit $\Delta\phi'$ parameter obtained in CD is higher than that given in FFA. The possible reason is that the integration of FFA from LIS to TOA is constrained by the characteristic curve, and the corresponding path of FFA is longer than that of CD for a same value of $\phi$. In other words, the adiabatic energy term of TPE ignored in CD is larger than that ignored in FFA. For both models, the largest $\Delta\phi'$ appears around 2014, which corresponds to a solar maximum. 

\begin{figure*}[htpb!]
    \centering
    \includegraphics[width=0.8\textwidth]{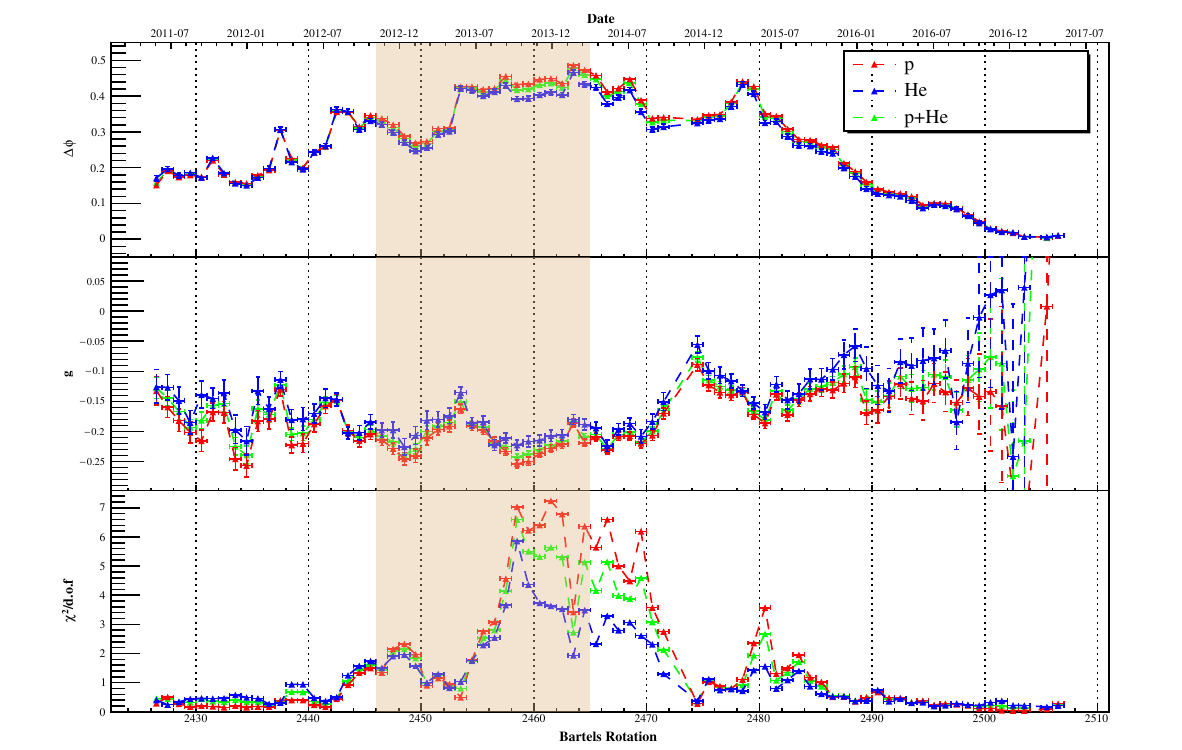}
    \caption{Top and middle panels: the rigidity-independent parameter $\Delta\phi'$ (=$\Delta\phi$) and $g$ estimated by using the p, He and He isotopes periodic data measured by AMS--02 for Kuhlen's EFFA. Bottom panel: the corresponding $\chi^2 / d.o.f$ at each BR. The shade area corresponds to the HMF polarity reversal period from $A < 0$ to $A>0$ (2012 November--2014 March).}
    \label{fig:effa-pdf}
\end{figure*}

\begin{figure*}[htpb!]
    \centering
    \includegraphics[width=0.45\textwidth]{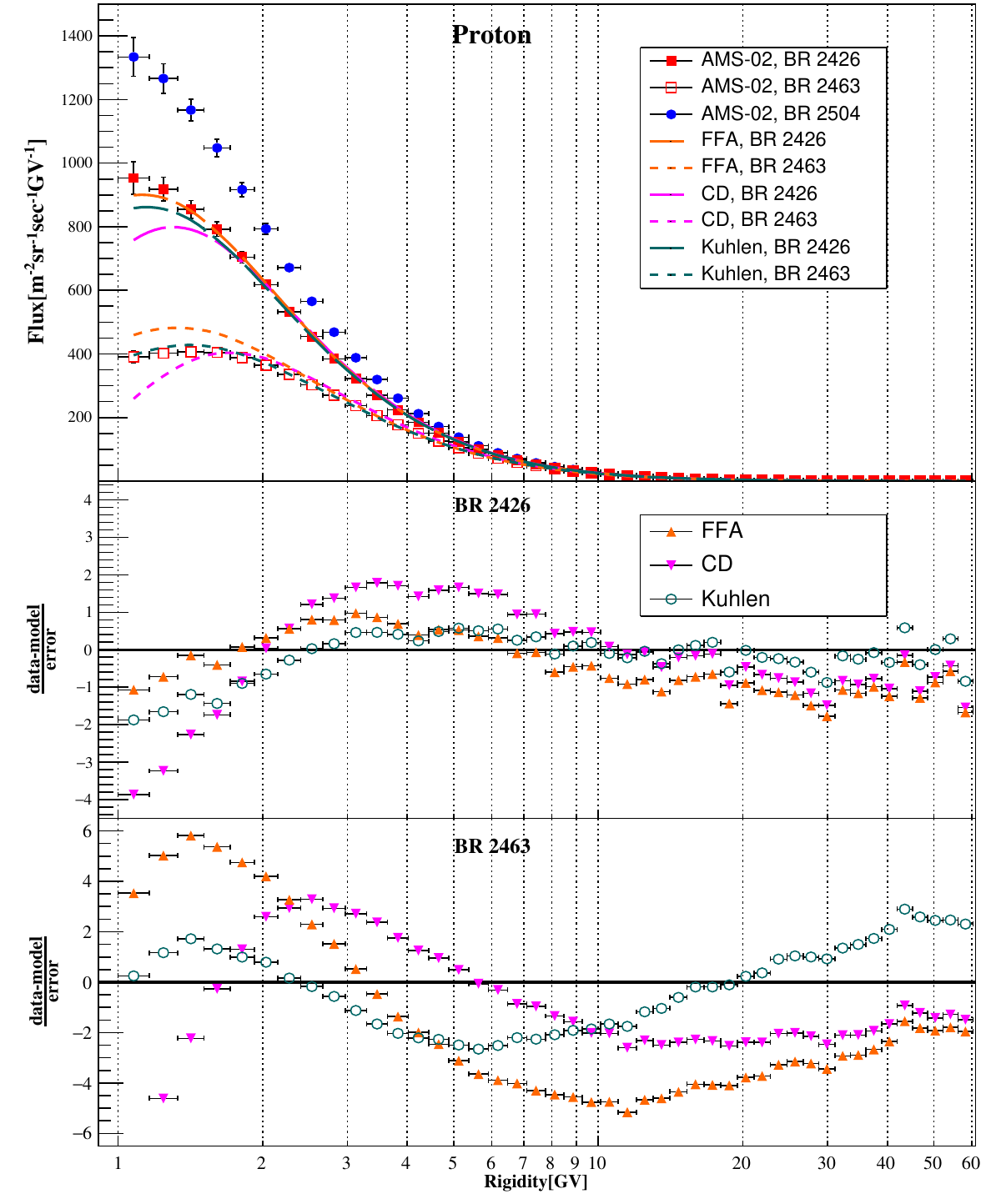}
    \includegraphics[width=0.45\textwidth]{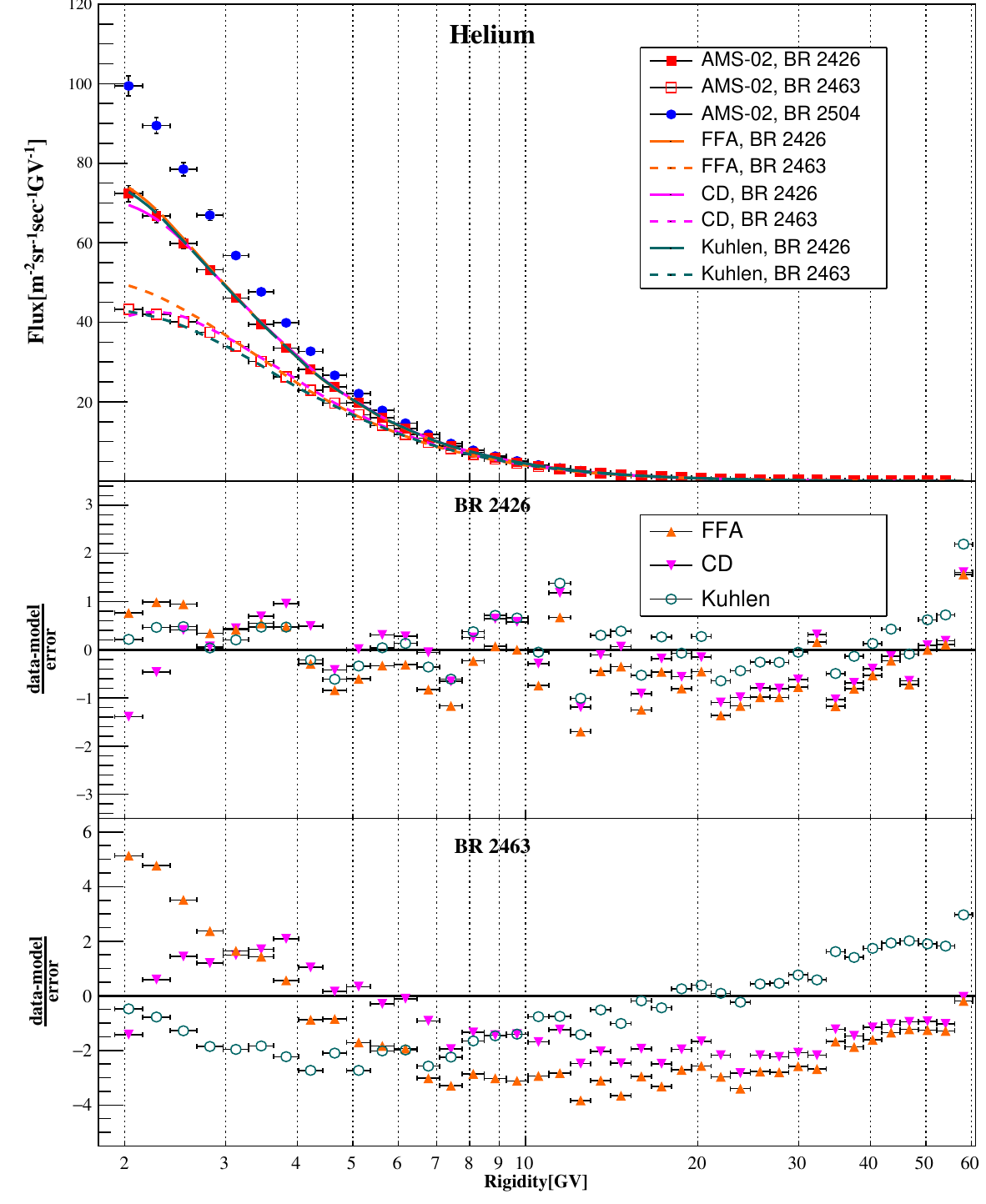}
    \caption{Top panels: The p and He fluxes at BR 2426 and BR 2504 expected from FFA, CD and Kuhlen's EFFA, in comparison with the AMS-02 measurements. Middle and bottom panels: the residuals of the model fittings to the p and He  spectra for BR 2426 and BR 2463.}
    \label{fig:results1}
\end{figure*}

During periods around 2011--2012 and 2016--2017, it is found that the $\chi^2/d.o.f$ values are close to 1. This infers that FFA and CD can generally reproduce data during the low solar activity periods. But during the periods with high $\Delta\phi'$, the $\chi^2/d.o.f$ values are much larger than 1. Especially, by using the p and He fluxes, the $\chi^2/d.o.f$ values can achieve more than 20. It means that, for both FFA and CD, an rigidity-independent $\Delta\phi'$ (or $\phi'$) does not agree well with the p and He data during these periods with intense solar activity. This phenomenon is further confirmed in Fig.~\ref{fig:results1}. At BR 2426 (May 5, 2011--June 10, 2011), a period after the solar minimum in 2009, the solar activity is not strong. As we can see, at this BR, both FFA and CD generally give consistent results with the p data above 2 GV and the He data at the whole rigidity range , except for an obvious discrepancy exist between CD and the p data below 2~GV. But the p and He fluxes predicted by both models at BR 2463 (February 7, 2014--March 6, 2014), which is in a polarity reversal in solar cycle 24, show significant disagreements with the data. These disagreements indicate that both models with a rigidity-independent $\phi'$ can not describe the solar modulation behavior well during the HMF polarity reversal period.

By using the $^{3}$He or $^{4}$He data, the calculated $\chi^2/d.o.f$ values at solar maximum are not that high. This may be due to the large errors existing in the $^{3}$He and $^{4}$He data, which infers that using the He isotope data alone is not enough to test the validity of solar modulation models. Therefore, to further investigate EFFA, we use only the p and He data to run the analysis. 

For EFFA, the time variations of $\Delta\phi$ and $g$ are shown in Fig.~\ref{fig:effa-pdf}. This model presents a similar tendency of $\Delta\phi'$ ($\phi'$) in terms of time with FFA and CD. The scaling factor $g$ does not show a clear variation with time. The minor fluctuations of $g$ indicate the intensity of drift effect does not vary greatly in different periods. According to the $\chi^2/d.o.f$ values given in Fig.~\ref{fig:effa-pdf}, it can be found that EFFA can better fit the data than FFA and CD. At all the periods, EFFA reduces the $\chi^2/d.o.f$ values by more than half. But EFFA still displays a poor performance during 2013--2014. This can also been seen in Fig.~\ref{fig:results1}. It seems that including a drift effect still cannot explain the solar modulation effect during the polarity reversal period. 

\subsection{Modulation with rigidity-dependent $\phi'$}
\begin{figure*}[htpb!]
    \centering
    \includegraphics[width=0.45\textwidth]{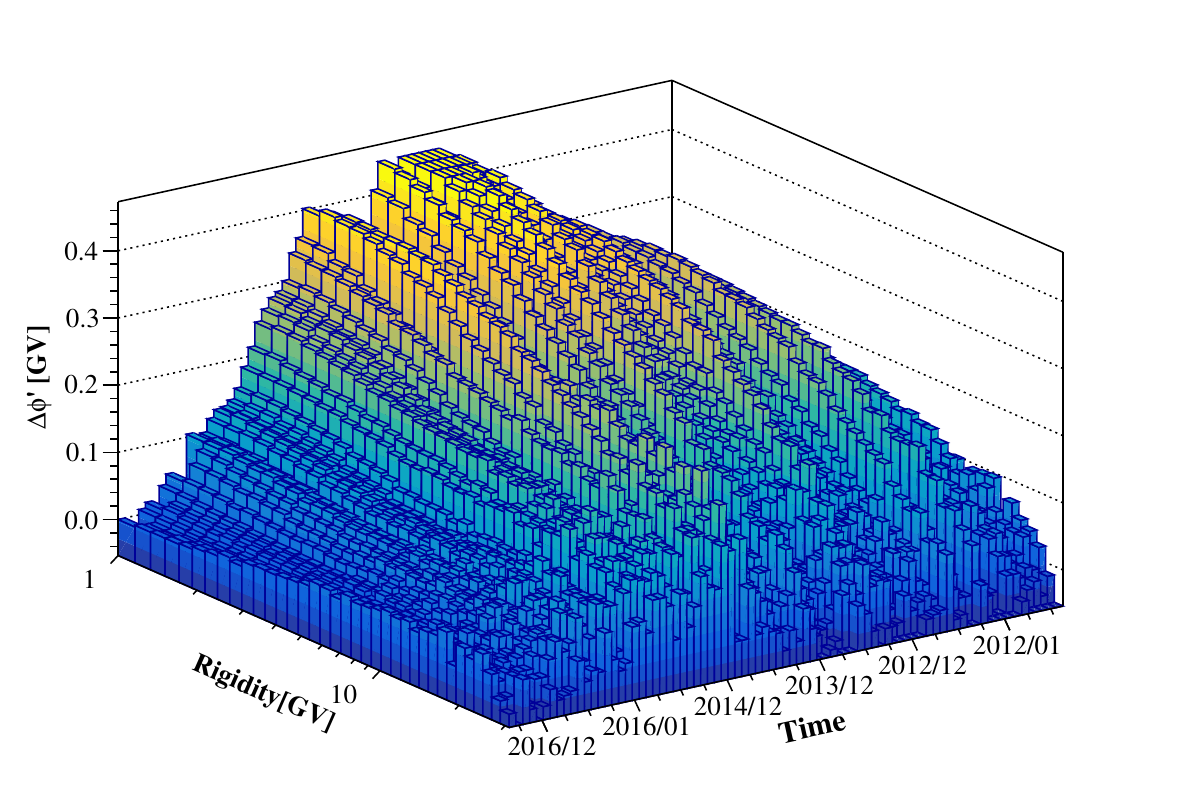}
    \includegraphics[width=0.45\textwidth]{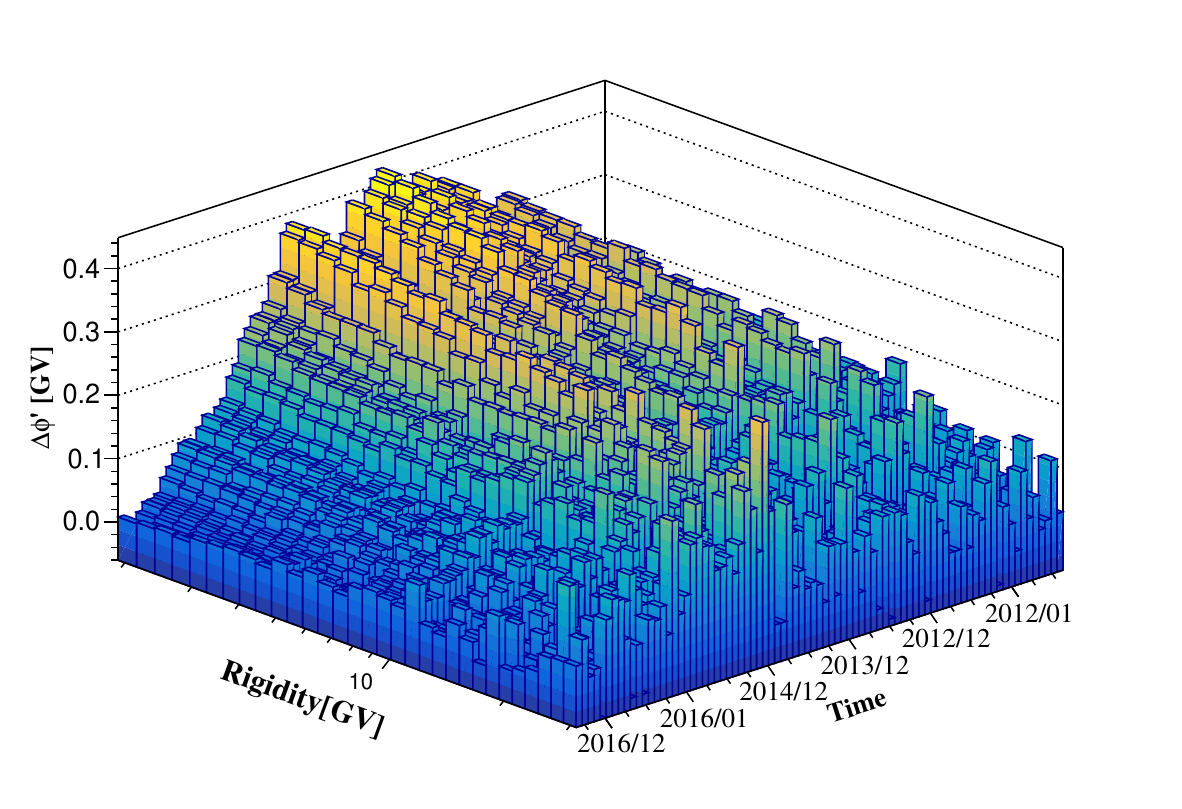}
    \caption{The variation of the best-fit parameter $\Delta\phi'$ with rigidity and time for p (left panel) and He (right panel).}
    \label{fig:ams02dphiprime}
\end{figure*}

Since a rigidity-independent $\phi'$ cannot accommodate the data well during the periods with intense solar activities, we only assume a constant $\phi'$ at solar minimum period $t_1$. But at other periods, we calculate $\Delta\phi'$ at each rigidity bin to study the change of $\Delta\phi'$ ($\phi'(t_2)$) with rigidity. The analysis is performed both for AMS-02 p and He data and the results for FFA are presented in Fig.~\ref{fig:ams02dphiprime}. Noted that in this figure the x-axis is represented on a logarithmic scale. 

\begin{figure}[htpb!]
    \centering
    \includegraphics[width=0.45\textwidth]{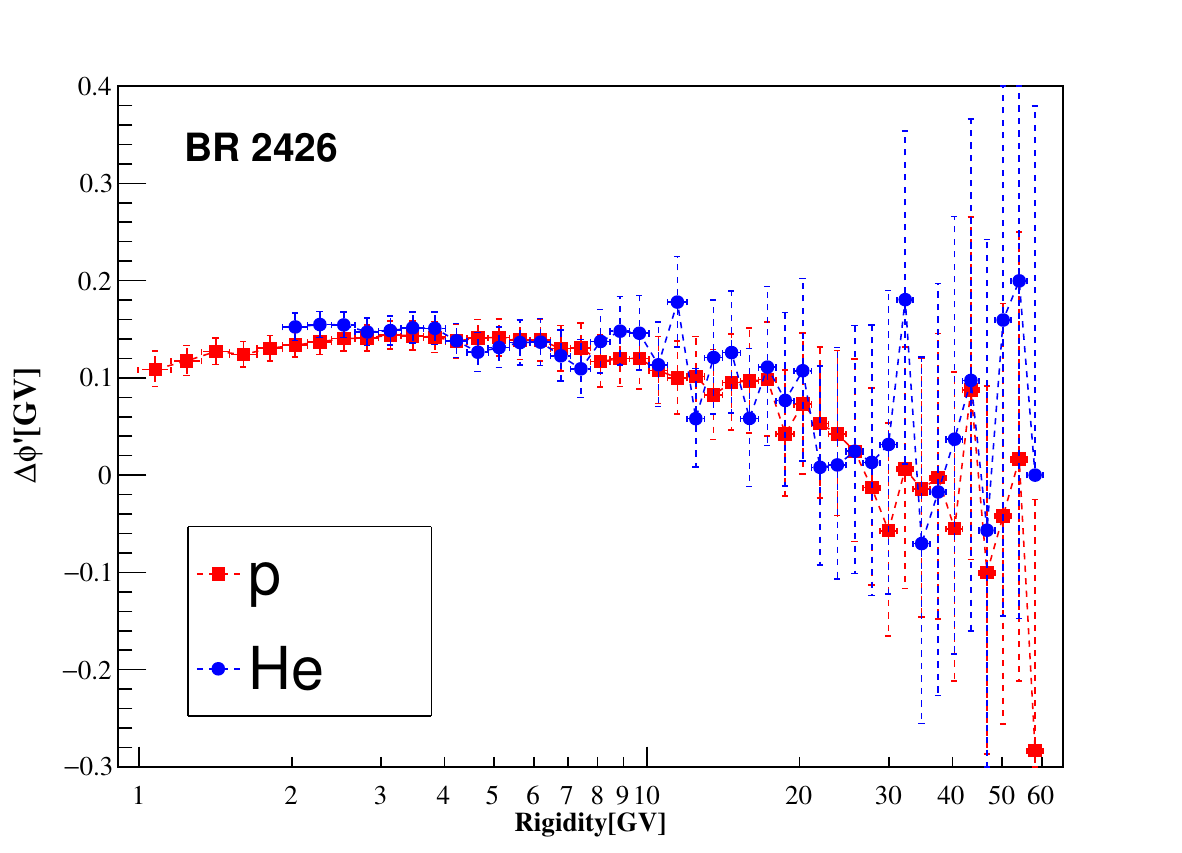}
    \includegraphics[width=0.45\textwidth]{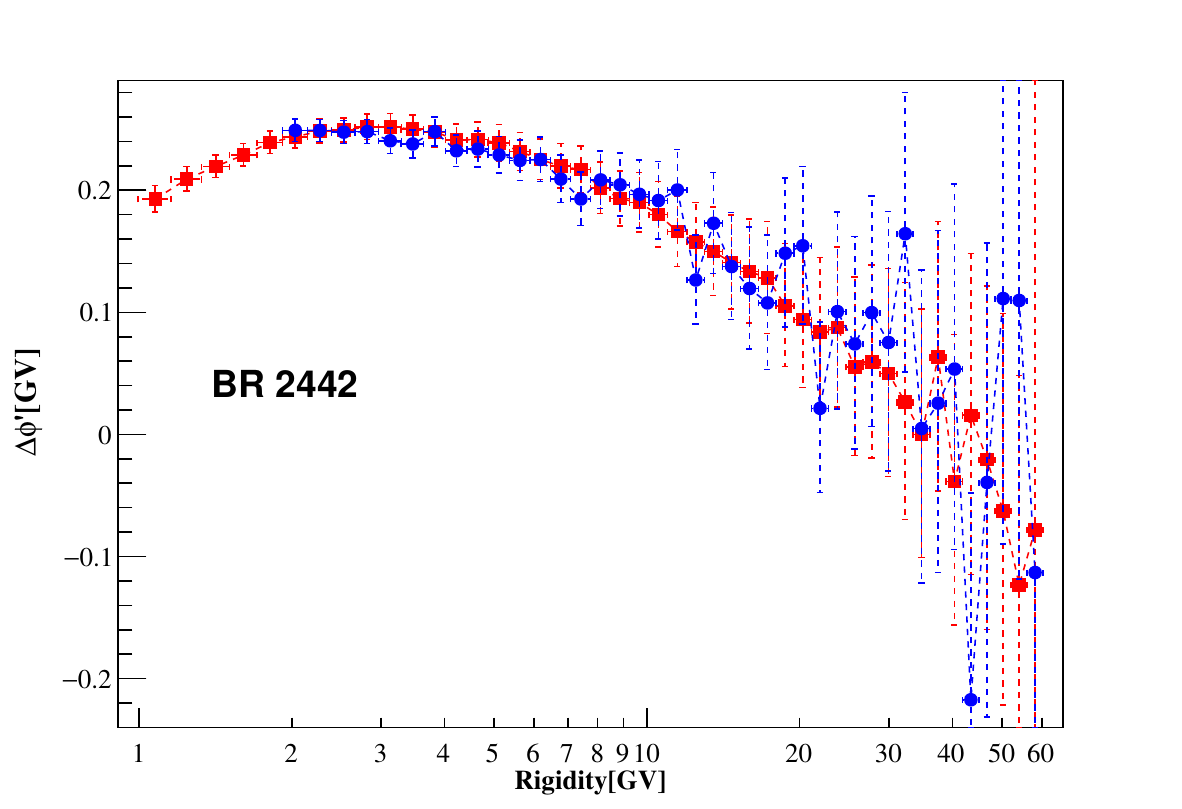}
    \includegraphics[width=0.45\textwidth]{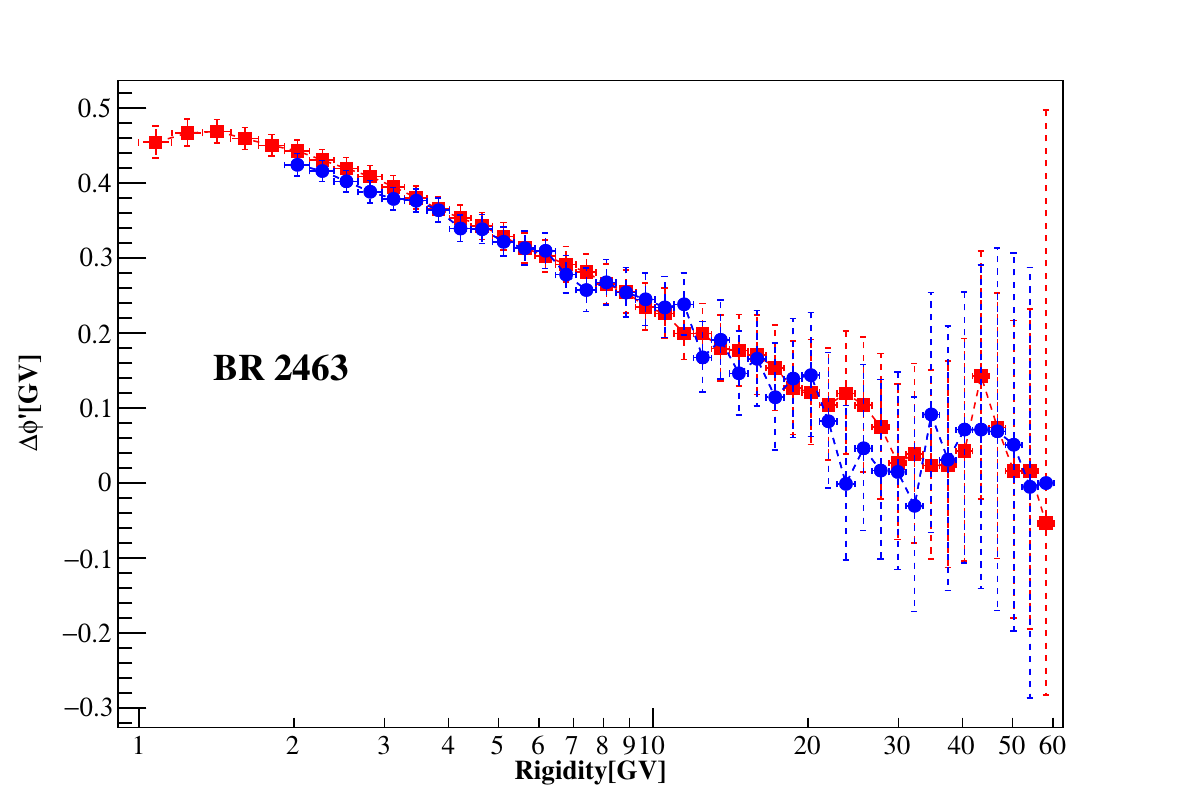}
    \caption{The parameter $\Delta\phi'$ as a function of rigidity in BR 2426, BR 2442, BR 2463, where BR 2426 is near the solar minimum, BR 2442 is one of the sharp dips observed in AMS02~\citep{amscollaborationTemporalStructuresElectron2023}, and BR 2463 is during the solar maximum period.}
    \label{fig:ams02deltaphiprime2}
\end{figure}

As the intensity of solar activity increases, the variation of $\Delta\phi'$ with rigidity becomes more and more significant. This further confirms the necessary to introduce a rigidity-dependent $\phi'$ during periods of high solar activity. We particularly show the relationships between $\Delta\phi'$ and rigidity in Fig.~\ref{fig:ams02deltaphiprime2} for BR 2426, BR 2442 and BR 2463. It can be found that for all BRs, the curves are very close to straight lines in linear-logarithmic (lin-log) coordinates. Here we include BR 2442 in the plot because this BR is related to the location of sharp dips in  the p and electron fluxes observed by AMS02~\citep{amscollaborationTemporalStructuresElectron2023}. At this BR, $\Delta\phi'$ has a slight downturn at very low rigidity. For CD and EFFA, $\Delta\phi'$ have similar relations with rigidity. Therefore, we assume that $\Delta\phi'$ has a lin-log relationship with rigidity, with the formula:
\begin{equation}
    \label{eq:linlogfit}
    \Delta\phi'_{\text{lin-log}} = \phi_0 + \phi_1 \ln\left( R / R_0\right).
\end{equation}
where $\phi_0$ is the normalization of $\Delta\phi'$ at $R_0 = 1$ GV, and $\phi_1$ is the slope of $\Delta\phi'$ with $\ln{R}$. They both vary with time.

\begin{figure*}[htpb!]
    \centering
    \includegraphics[width=0.45\textwidth]{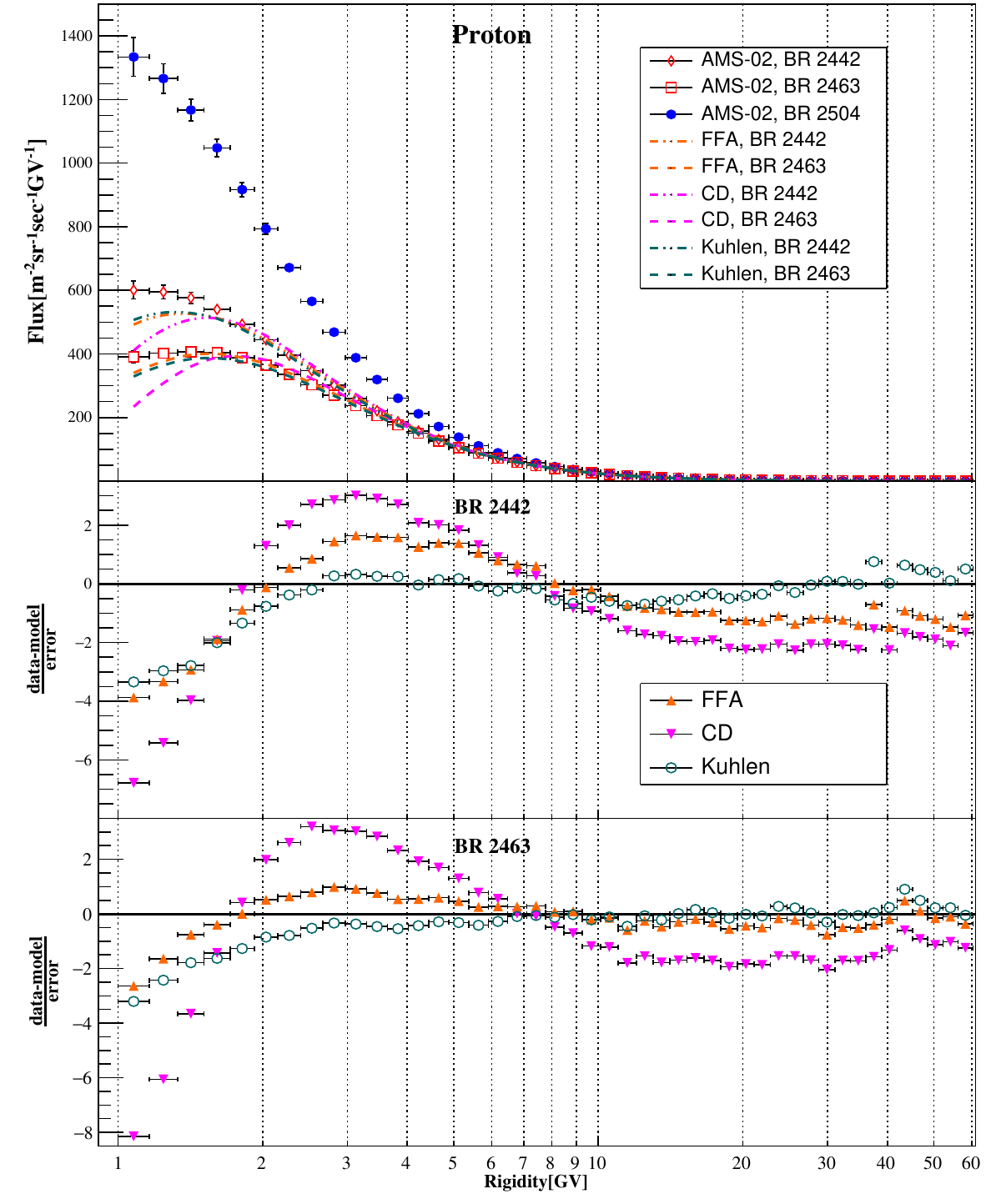}
    \includegraphics[width=0.45\textwidth]{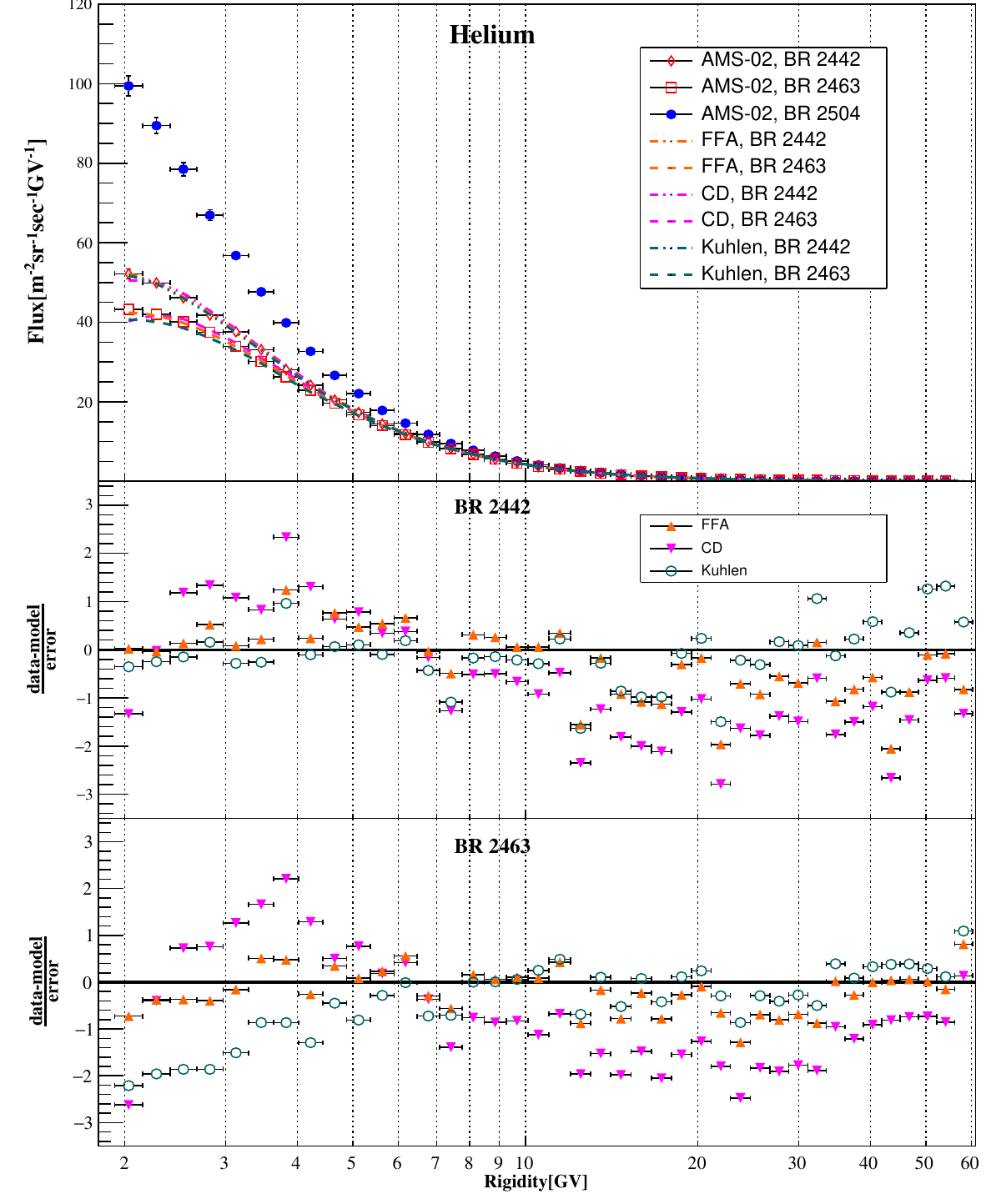}
    \caption{Top panels: The p and He fluxes at BR 2426 and BR 2504 expected from the modified FFA, CD and Kuhlen's EFFA by introducing a linear-logarithm rigidity-dependent $\Delta\phi'$, in comparison with the AMS-02 measurements. Middle and bottom panels: the residuals of the model fittings to the p and He  spectra for BR 2442 and BR 2463.}
    \label{fig:fitresults1} 
\end{figure*}

We adopt this lin-log formula of $\Delta\phi'$ in FFA, CD and EFFA. In this case, the free parameters include $\phi_0$, $\phi_1$ for FFA and CD, and $\phi_0$, $\phi_1$, $g$ for EFFA. The predicted p and He flux are compared with the data measured at BR 2442 and BR 2463, as shown in Fig. \ref{fig:fitresults1}. At BR 2442 and 2463, the modified FFA can generally reproduce the p and He data at most rigidity range. It only gives slightly lower predictions than the p data below 2~GV. This might because $\Delta\phi_{\text{lin-log}}$ does not give a perfect description of the p flux at very low rigidity, as exhibited in Fig.~\ref{fig:ams02deltaphiprime2}. Nevertheless, the agreement between the modified FFA and the data is highly increased compared with the conventional FFA. This is also the case for modified EFFA. But the improvement of the modified CD is limited compared with the conventional CD. It give worse goodness of fit than the modified FFA and EFFA. 

\begin{table}[htpb!]
    \centering
    \caption{The $\chi^2/d.o.f$ results for different analytical models at BR 2426, BR 2442 and BR2463 based on the analysis of p or (and) He data. Here $\phi'=\phi$ assumes a rigidity-independent $\phi'$, and $\phi'_{\text{lin-log}}$ assumes a linear-logarithm rigidity-dependent $\phi'$.}
    \label{tab:model-comparison}
\begin{tabular}{llcrrr}
\hline
BR   &Model  & $\phi'$ &  $\chi^2_{\text{p}} / d.o.f$&$\chi^2_{\text{He} } / d.o.f$&  $\chi^2_{\text{p+He} }/ d.o.f$   \\
\hline
2426 &FFA   & $\phi$                   &  35.1/44  & 6.8/39   & 65.0/84 \\
2426 &FFA   & $\phi'_{\text{lin-log}}$ &  30.0/43  & 13.4/38  & 53.9/83 \\
2426 &CD    & $\phi$                   &  75.3/44  & 18.5/39  & 108.3/84 \\
2426 &CD    & $\phi'_{\text{lin-log}}$ &  54.3/43  & 18.5/38  & 86.7/83 \\
2426 &EFFA& $\phi$                   &  12.3/43  & 15.2/38  & 37.6/83 \\
2426 &EFFA& $\phi'_{\text{lin-log}}$ &  5.8/42   & 12.9/37  & 33.4/82 \\
      
2442 &FFA   & $\phi$                   &  178.4/44 & 109.5/39 & 289.2/84\\
2442 &FFA   & $\phi'_{\text{lin-log}}$ &  84.7/43  & 20.0/38  & 110.1/83\\
2442 &CD    & $\phi$                   &  289.8/44 & 60.6/39 & 354.2/84\\
2442 &CD    & $\phi'_{\text{lin-log}}$ &  238.2/43 & 59.7/38  & 321.8/83\\
2442 &EFFA& $\phi$                   &  18.9/43  & 19.1/38  & 39.4/83 \\
2442 &EFFA& $\phi'_{\text{lin-log}}$ &  18.8/42  & 12.3/37  & 38.1/82 \\
      
2463 &FFA   & $\phi$                   &  585.2/44 & 283.8/39 & 906.8/84\\
2463 &FFA   & $\phi'_{\text{lin-log}}$ &  18.7/43  & 9.4/38   & 29.2/83 \\
2463 &CD    & $\phi$                   &  237.6/44 & 105.1/39 & 344.7/84\\
2463 &CD    & $\phi'_{\text{lin-log}}$ &  233.2/43 & 62.1/38  & 314.3/83\\
2463 &EFFA& $\phi$                   &  146.8/43 & 73.6/38 & 226.4/83\\
2463 &EFFA& $\phi'_{\text{lin-log}}$ &  6.3/42   & 5.6/37   & 13.7/82 \\
\hline
\end{tabular}
\end{table}

The $\chi^2/d.o.f$ results of different models are summarized in Table \ref{tab:model-comparison}. As we can see, CD doesn't fit well with all the data. For BR 2426, both the FFA and EFFA with a rigidity-dependent or rigidity-independent $\phi'$ agrees well with the p and He data. For BR 2442, the conventional FFA does not accommodate the data. The modified FFA improves the goodness-of-fit but still yield a $\chi^2/d.o.f$ close to 2. By including a drift effect, both conventional EFFA and modified EFFA can reproduce the p and He data at BR 2442. It suggests that the flux dips observed in AMS02 data may be associated with the drift effect. For BR 2463, the conventional FFA and EFFA have large disagreements with the data. But by adopting a rigidity-dependent $\phi'$, both models can explain the data well. For solar minimum and maximum phases, it is difficult do judge whether the drift effect needs to be introduced to interpret the data.

\subsection{Compared with other modified FFA models}

We compared our lin-log FFA and lin-log EFFA with other modified FFA models. One is Cholis' model~\cite{cholisPredictiveAnalyticModel2016,cholisConstrainingChargesignRigiditydependence2022}. Instead of adding a drift term in the relationship between $J(r_{TOA},t_2)$ and $J(r_{TOA},t_1)$, they incorporated the drift term in $\phi$. Based on their work, $\Delta\phi'$ can be written as
\begin{equation}\label{eq:cholis}
    \Delta\phi'_{\text{Cholis}} = \phi_0 + \phi_1 \left( \frac{1+\left( R / R_0 \right) ^2}{\beta\left( R / R_0 \right) ^3} \right)
.\end{equation} 
The other is Shen's model presented in \cite{shenSolarModulationGalactic2021}. In that paper, the authors attributed the variation of $\phi$ with energy to the behavior of diffusion coefficient. They used a double power-law empirical formula to describe $\phi$. In Shen's model $\Delta\phi'$ can be written as
 \begin{equation}\label{eq:shen}
     \Delta\phi'_{\text{Shen}} = \phi_0 \beta^{-1} \left( \frac{E}{E_b} \right) ^{\phi_1} \left[ 1+\left( \frac{E}{E_{b_1}} \right)^{b1}  \right] ^{b_2}
,\end{equation}
where $E_b = 1$ GeV, $\phi_0$ is a scaling factor in unit GV. The rest of the parameters are dimensionless. Both $\phi_0$ and $\phi_1$ vary with time, while $E_{b_1}$, $b_1$ and $b_2$ are time-independent parameters. It is worth to noticed that the estimations of all the parameters in Shen's model are adjustable to perform a good agreements with the data. This could result in overfitting and instability of the parameters \cite{songNumericalStudySolar2021,tomassettiDataDrivenAnalysis2023}. 

Both Cholis' and Shen's models include a $\beta$ term from the diffusion coefficient into $\phi'$. The relationship between $\beta$ and rigidity shows that $\beta$ is a function of $A/Z$. Thus for different particles, the same values of $\phi_0$ and $\phi_1$ may lead to different values of $\Delta\phi'_{\text{Shen}}$ (or $\Delta\phi'_{\text{Cholis}}$) for a given rigidity. This difference is slight in Cholis' model since $\beta$ only exist in $\phi_1$ term of $\Delta\phi'_{\text{Cholis}}$. But from Eq.~\eqref{eq:modifiedphi}, we can see that a $\beta$ term is unnecessary to be introduced in $\phi'$. The inclusion of $\beta$ may be lack of rigorous theoretical basis. 

\begin{figure*}[htpb!]
    \centering
    \includegraphics[width=0.45\textwidth]{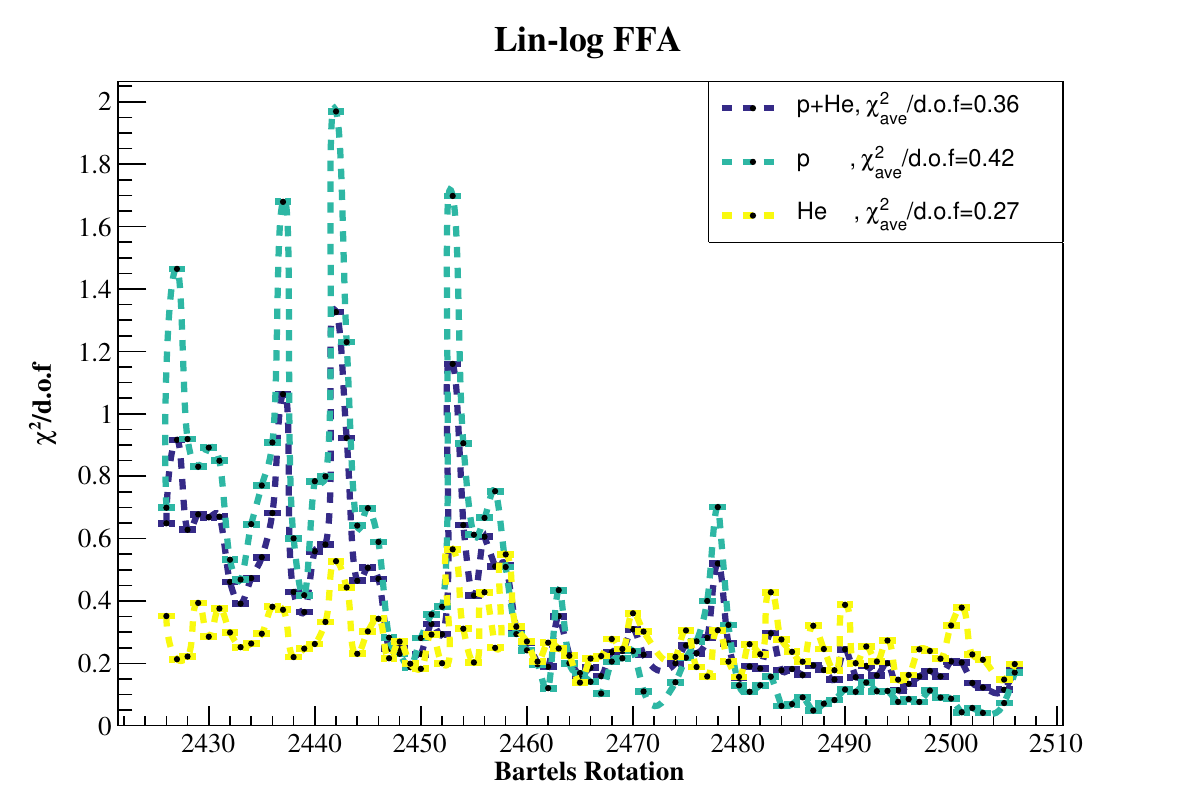}
    \includegraphics[width=0.45\textwidth]{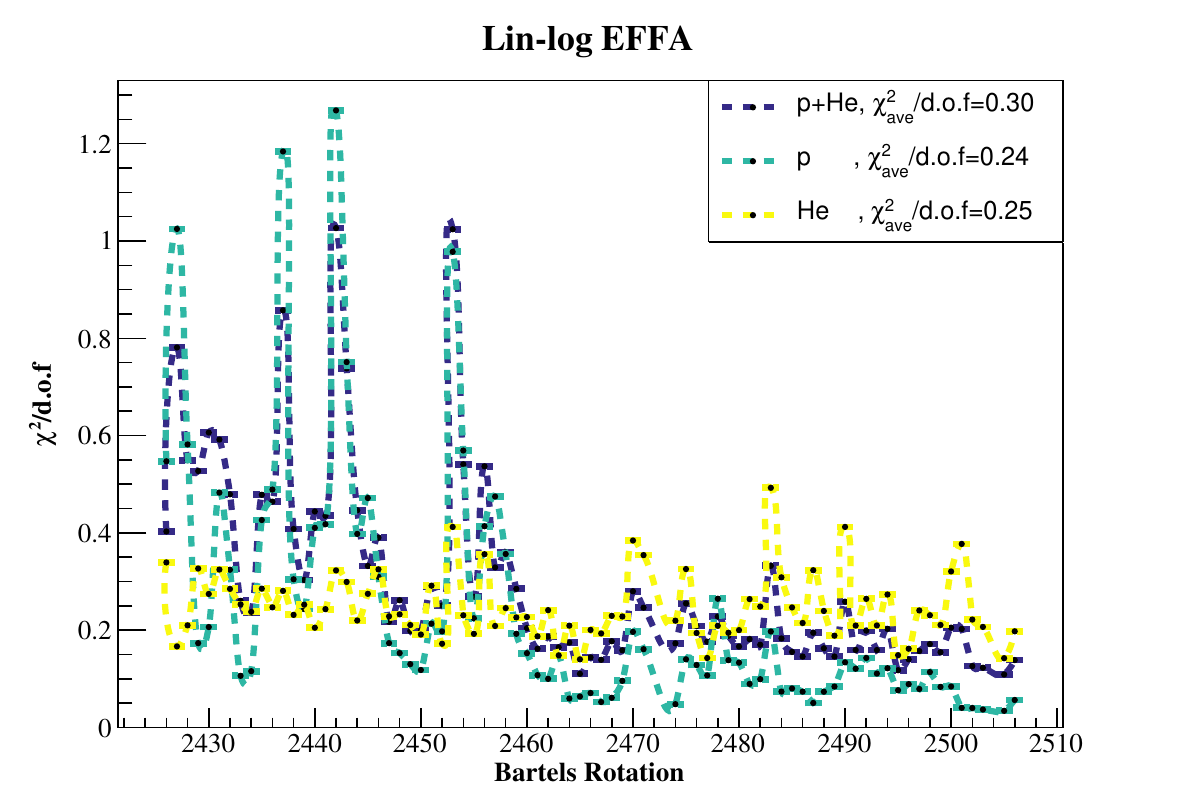}
    \includegraphics[width=0.45\textwidth]{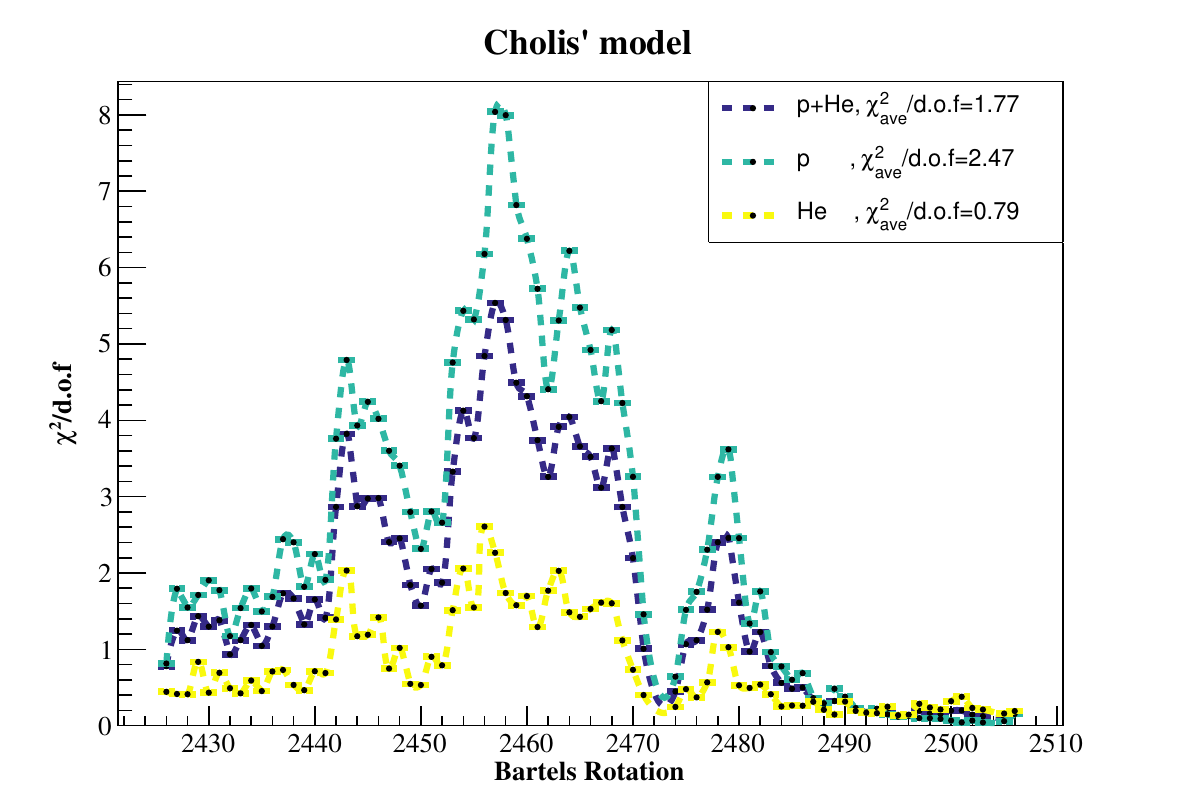}
    \includegraphics[width=0.45\textwidth]{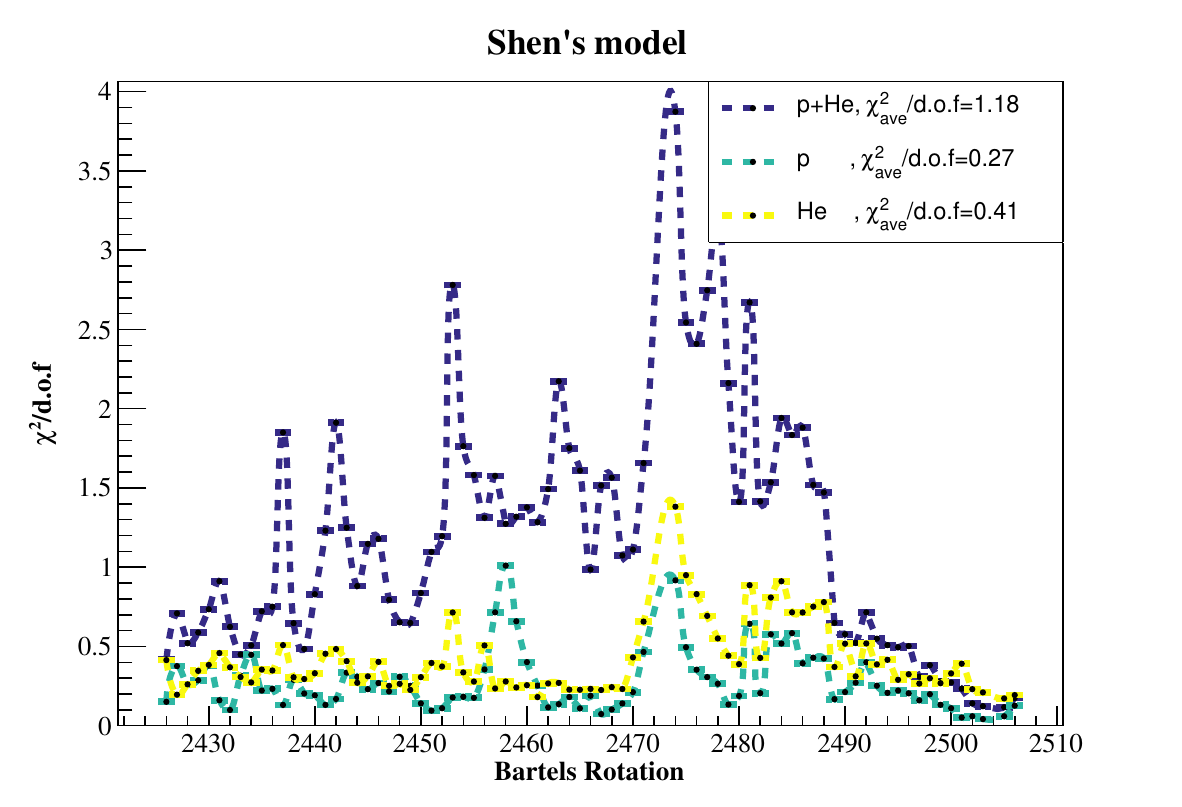}
    \caption{The results of $\chi^2 / d.o.f$ over time obtained by fitting the p or (and) He data for the modified FFA, the modified EFFA, Cholis' model and Shen's model. }
    \label{fig:modelchi2}
\end{figure*}

Above models with a rigidity-dependent $\Delta\phi'$ all contain two free parameters $\phi_0$ and $\phi_1$. Other parameters are nuisances. The $\chi^2$ minimization results of the lin-log FFA, the lin-log EFFA, Cholis' model and Shen's model are shown in Fig.~\ref{fig:modelchi2}, respectively.

It can be found that the lin-log FFA give excellent goodness-of-fit for most periods. There are only a few BRs at which the values of $\chi^2 / d.o.f > 1$. Peaks of $\chi^2$ appears at BR 2437, BR 2442, BR 2453 and BR 2478. All these BRs happens only in $A>0$ stage and corresponds to the sharp dips in AMS-02 p and electron fluxes. Notably, at these BRs, the lin-log EFFA agrees better with the AMS02 p and He data. It indicates that these solar transients on timescale of BRs maybe related with the drift effect. 

For Cholis' model, there are much more BRs corresponding to $\chi^2 / d.o.f > 1$, especially by fitting the p or p+He data. The largest $\chi^2$ values exhibit during the solar reversal phase, which means Cholis' model is particularly poor to simulate the solar modulation during those stages. In this model, the $\chi^2$ distribution over time is similar with that in conventional EFFA. This suggests that the introduction of a drift effect is not sufficient to explain the variation of $\phi'$ with rigidity. 

Shen's model could obtain good agreements with either p or He data. But when we combine p and He data to do the analysis, the resulted $\chi^2/d.o.f$ values are large. The reason is that the estimated $\phi_0$ and $\phi_1$ deviate significantly between p and He. It reveals that Shen's model does not give consistent descriptions on p and He. 

\section{Conclusion and Discussion}
In this paper, we take into account three analytic solar modulation models: FFA, CD, and EFFA. To investigate these models, the Non-LIS method is employed to eliminate the impact of CR LIS spectra. The traditional potential parameter $\phi$ in FFA, CD and EFFA is a rigidity-independent parameter. However, since the radial diffusion coefficient may not be proportional to rigidity, the parameter $\Phi$ could be rigidity-dependent. Therefore, we introduce an alternative parameter $\phi' =\frac{R}{ k_2(R)}\phi$ to revisit these models. By using $\Delta \phi'=\phi'(t_2)-\phi'(t_1)$, we can determine the CR flux at time $t_2$ based on the observed CR flux at time $t_1$. Then we use the $\chi^2$ minimization analysis to estimate the best-fit $\Delta \phi'$.  

First, it is found that the conventional FFA and EFFA with a rigidity-independent $\phi'$ can describe the data well around solar minimum. But these models do not agree well with the data at HMF polarity reversal periods. Therefore, it is reasonable to assume a constant $\phi'$ near solar minimum, but consider a rigidity-dependent $\phi'$ (or $\Delta\phi'$) for other periods. By calculating $\Delta\phi'$ at different rigidity ranges, the results show that $\Delta\phi'$ is not a constant but seems have a lin-log relationship with rigidity. By incorporating this lin-log formula of $\Delta\phi'$ into FFA and EFFA models, we find that they can satisfactorily describe the data during the HMF polarity reversal stage. The CD models, no matter the conventional one or the modified one, cannot explain the data well. It infers that the ignored adiabatic term in CD plays a relatively important role in modulation, which could have a significant rigidity dependence.

The effect of drift may be important to explain the modulation during those solar transients detected by AMS02, since during those stages, the conventional and modified EFFA models can fit the date better than other models. The hysteresis-like loops (coinciding with the sharp dip times) between the proton and electron fluxes~\cite{amscollaborationTemporalStructuresElectron2023} or between the proton and antiproton fluxes~\cite{aslamModulationCosmicRay2023} display a charge-sign-dependent solar modulation. This may be related to the fact that particles with opposite charge signs have different patterns of drift effects. Nevertheless, the variation of $\phi'$ with rigidity is not mainly due to the drift effect, since Cholis' model have a worse performance than our lin-log FFA model. This suggests that the rigidity-dependence of the parameter $\phi'$ mainly originates from the rigidity-dependence of diffusion coefficient. The lin-log FFA model is also better than Shen's model, in which they consider a double power-law $\phi'$. The specific form of $\phi'$ is important for understanding of HMF fluctuations during the HMF polarity reversal periods. 

A recent study by simultaneous scanning on the solar modulation parameter and other CR acceleration and propagation parameters, has suggested that the conventional FFA can describe well the CR spectra measured by AMS02 and Voyager-1 integrated over the entire detection period~\citep{silverTestingCosmicRay2024}. However, a rigidity-dependent $\phi'$ may challenge our traditional understanding on CR acceleration and propagation mechanisms.

It should be noted that these analytic models are based on a series of assumptions. The dependence of modulation on A/Z is not studied in this work. But we find that during the sharp dips periods, the deviations of the He data from lin-log FFA are less significantly than those of the p data. This will be further studied in our future work. The solar modulation model proposed in our work enables us to place effective constraints on the CR source and propagation models. This allows for a reliable calculation on the CR LIS spectra. Some other studies have shown that it is also possible to derive the LIS spectra from synchrotron and gamma-ray observations without any assumption on solar modulation~\cite{strongVizieROnlineData2011,ackermannFermiLATObservationsDiffuse2012,orlandoGalacticSynchrotronEmission2013,orlandoImprintsCosmicRays2018}. In our future work, we will further compare them for a better understanding on cosmic ray behaviors in Galaxy.

\begin{acknowledgments}
Thanks for Ilias Cholis, Claudio Corti and R.A. Caballero-Lopez for very helpful discussions. This work is supported by the Joint Funds of the National Natural Science Foundation of China (Grant No. U1738130). The use of the high-performance computing platform of China University of Geosciences is gratefully acknowledged.
\end{acknowledgments}
\bibliography{ref}{}

\end{document}